# Entropy Engineered Middle-In Synthesis of Dual Single-Atom Compounds for Nitrate Reduction Reaction

Yao Hu,1§ Haihui Lan,2§ Junjun He,1 Wenjing Fang,1 Wen-Da Zhang,1 Shuanglong Lu,1 Fang Duan,1 Mingliang Du1*

1 Key Laboratory of Synthetic and Biological Colloids, Ministry of Education, School of Chemical and Material Engineering, Jiangnan University, Wuxi, Jiangsu 214122, P. R. China.
2 Department of Chemistry, Massachusetts Institute of Technology, Cambridge, Massachusetts 02139, United States.

**ABSTRACT:** Despite the immense potential of Dual Single-Atom Compounds (DSACs), the challenges in their synthesis process, including complexity, stability, purity, and scalability, remain primary concerns in current research. Here, we present a general strategy, termed "Entropy-Engineered Middle-In Synthesis of Dual Single-Atom Compounds" (EEMIS-DSAC), which is meticulously crafted to produce a diverse range of DSACs, effectively addressing the aforementioned issues. Our strategy integrates the advantages of both bottom-up and top-down paradigms, proposing a new insight to optimize the catalyst structure. The as-fabricated DSACs exhibited excellent activity and stability in the nitrate reduction reaction ($NO_3RR$). In a significant advancement, our prototypical CuNi DSACs demonstrated outstanding performance under conditions reminiscent of industrial wastewater. Specifically, under a $NO_3^-$ concentration of 2000 ppm, it yielded a Faradaic efficiency (FE) for $NH_3$ of 96.97 %, coupled with a mass productivity of 131.47 mg h$^{-1}$ mg$^{-1}$ and an area productivity of 10.06 mg h$^{-1}$ cm$^{-2}$. Impressively, even under a heightened $NO_3^-$ concentration of 0.5 M, the FE for $NH_3$ peaked at 90.61 %, with a mass productivity reaching 1024.50 mg h$^{-1}$ mg$^{-1}$ and an area productivity of 78.41 mg h$^{-1}$ cm$^{-2}$. This work underpins the potential of the EEMIS-DSAC approach, signaling a promising frontier for high-performing DSACs.

______________________________________________________

## INTRODUCTION

Dual Single-Atom Compounds (DSACs) have emerged as a focal point in the field of catalysis, drawing widespread attention due to their unique properties and potential application value[1]. Characterized by the uniform dispersion of two distinct metal atoms on a singular support, this advanced catalytic system not only inherits the high activity and selectivity of Single-Atom Catalysts (SACs)[2,3] but also amplifies catalytic performance through the synergistic interaction between the two metal atoms[4-7]. However, despite the immense potential of DSACs, there are key challenges to overcome in their synthesis. Firstly, current synthesis methods are often complex and costly, limiting their broad adoption in industrial applications[8-10]. Secondly, the stability of DSACs[11-13], especially the phenomena of deactivation during catalytic reactions, has become a major barrier to their long-term stable application. Additionally, the purity of the catalyst is an issue that cannot be overlooked[14], as any impurities or cluster structures can adversely affect its performance and activity. Lastly, most of the current synthesis methods are tailored to specific metals and substrate materials, greatly restricting the scalability and widespread application of DSACs[15,16]. Therefore, to harness the advantages of DSACs to the fullest, it's imperative to address these challenges, refine synthesis techniques, and develop more universal and efficient synthesis methods[17].

In the realm of SACs synthesis, two predominant strategies—top-down and bottom-up—each come with an array of inherent challenges and limitations[15,18-21]. The top-down approach typically starts with larger metal structures, such as nanoparticles or nanoclusters, and employs methods like electrochemical etching or thermal treatment to meticulously break them down into single atoms[22]. A primary challenge of this method is the need for precise control to prevent incomplete or excessive decomposition, which might lead to unstable intermediates, thereby affecting the stability and activity of the catalysts[23]. On the other hand, the bottom-up strategy begins with metal ions or organometallic compounds and builds up to the single-atom level through chemical or physical methods[24]. Drawbacks of this approach include the necessity for multi-step procedures and intricate control, and it might involve the use of potentially toxic, expensive, or environmentally unfriendly chemical agents[25]. In essence, both strategies grapple with inherent challenges related to material stability, product purity, environmental impact, temporal efficiency and scalability.

Herein, we present a novel and universal approach termed "Entropy-Engineered Middle-In Synthesis of Dual Single-Atom Compounds" (EEMIS-DSAC). This methodology proficiently mitigates challenges associated with intricacy, robustness, purity, and scalability, facilitating the direct fabrication of an extensive library of DSACs. We integrate the concept of entropy into SACs[26]. As the system's configurational entropy increases due to the atomic dispersion of metal atoms[27], this elevation in entropy might serve as the primary driving force for the formation of SACs. Situated at the nexus between the bottom-up and top-down paradigms, the EEMIS-DSAC strategy seamlessly integrates the strengths of both techniques. This "Middle-In" approach places special emphasis on the pivotal role of the "middle" phase. Through a bottom-up carbothermal reduction, we adeptly transformed metal salts into nanoparticles. Then, utilizing a top-down chemical vapor deposition technique, we meticulously achieved the transition from these nanoparticles to single-atom structures. Guided predominantly by the system's entropy increase, these two strategies collaboratively operate throughout the synthesis process, ensuring unparalleled atomic precision. Consequently, it provides a promising solution to various challenges encountered in the realm of catalysis[28]. As a proof-of-concept, the performance of the synthesized catalysts was evaluated through nitrate reduction reactions ($NO_3RR$), demonstrating their potential as electrocatalysts. Remarkably, the prototypical CuNi DSACs, under conditions with a $NO_3^-$ concentration of 2000 ppm(32.3mM), relevant to industrial wastewater, achieved a maximum FE for $NH_3$ of 96.97±0.11%, a mass productivity of 131.47±0.14 mg h$^{-1}$ mg$^{-1}$, and an area productivity of 10.06±0.01 mg h$^{-1}$ cm$^{-2}$(Catalyst unit area loading 0.0765 mg cm$^{-2}$). Under conditions with a NO$_3^-$ concentration of 0.1M, we achieved a maximum FE for NH$_3$ of 94.09±0.52%, a mass productivity of 239.10±0.60 mg h$^{-1}$ mg$^{-1}$, and an area productivity of 18.30±0.05 mg h$^{-1}$ cm$^{-2}$. Under conditions with a NO$_3^-$ concentration of 0.25M, we achieved a maximum FE for NH$_3$ of 94.56±0.46%, a mass productivity of 432.76±2.47 mg h$^{-1}$ mg$^{-1}$, and an area productivity of 33.12±0.19 mg h$^{-1}$ cm$^{-2}$. Under conditions with a NO$_3^-$ concentration of 0.5 M, we achieved a maximum FE for NH$_3$ of 90.61±0.16%, a mass productivity of 1024.50±0.99 mg h$^{-1}$ mg$^{-1}$, and an area productivity of 78.41±0.08 mg h$^{-1}$ cm$^{-2}$. Under conditions with a NO$_3^-$ concentration of 1M, we achieved a maximum FE for NH$_3$ of 91.71±0.66%, a mass productivity of 1023.88±7.35 mg h$^{-1}$ mg$^{-1}$, and an area productivity of 78.36±0.56 mg h$^{-1}$ cm$^{-2}$.

______________________________________________

## RESULTS AND DISCUSSION

### 1 Entropy Engineered Middle-In Strategy for Dual Single-Atom Compounds Synthesis

As illustrated in **Figure 1**a, EEMIS-DSAC encompasses a series of meticulously designed steps. Firstly, utilizing the electrospinning technique, we successfully fabricated a unique "sandwich" structured nanofiber membrane. The central layer of this membrane is composed of polyacrylonitrile (PAN) fibers doped with metal salts, while the outer layers on both sides are constructed entirely of pure PAN fibers. During the early stages of pre-oxidation, due to the exothermic reaction, PAN molecular chains undergo rupture, leading to the formation of a macroporous defect structure. Concurrently, the metal salts are progressively transformed into their corresponding metal oxides[29, 30].

Under an argon-protected environment, as the temperature gradually rises, the pre-oxidized electrospun fibers commence carbonization[31], accompanied by a carbothermal reduction process primarily driven by entropy increase[32]. Owing to the slow diffusion rate of solid reactants, heating is required to facilitate the carbothermal reduction. In the bottom-up approach, the solid metal oxides in the middle layer react with carbon, resulting in the formation of new metals and gaseous CO. The newly formed metals likely exist in three morphologies: 1) atomically dispersed, akin to SACs; 2) forming monolayer two-dimensional clusters; 3) through surface diffusion, monolayer clusters further evolve into three-dimensional crystalline particles, which may migrate and coalesce to form larger grains[33]. This stage can be considered a pivotal "middle" phase. Upon detailed characterization of the catalyst fibers at the intermediate stage, we distinctly identified metal nanoparticles of 1-2 nm in size through TEM analysis (Figure S1, Supporting Information). Additionally, XRD analysis (Figure S2, Supporting Information) unequivocally revealed the characteristic diffraction peaks of the CuNi alloy. Utilizing the Scherrer formula, we determined the average crystallite size to be approximately 2 nm. The findings are in strong alignment with our earlier research outcomes, providing compelling validation for the bottom-up carbothermal reduction strategy.

For further optimization, we elevated the temperature and employed chemical vapor deposition (CVD) technology, adopting a top-down strategy for fine-tuning. Upon reaching the evaporation temperature of metal atoms, the three metal morphologies revert to atomic form and initiate atomic diffusion. Given that the metal atom concentration in the middle layer is significantly higher than in the adjacent layers, according to Fick's diffusion law, metal atoms diffuse from high-concentration regions to low-concentration ones until concentration equilibrium is achieved[34, 35]. Notably, the entire diffusion process transitions from three-dimensional particles to two-dimensional clusters, and ultimately to individual atoms, illustrating a distinct top-down progression. Especially within the intermediate layer, it is anticipated that there will be a substantial presence of graphitic carbon defects inside the fibers. This assertion is corroborated by TEM analysis (Figure S3, Supporting Information) when compared to pure carbon fibers, as depicted in Figure S4, Supporting Information. In a high-temperature environment, the uniform dispersion of metal atoms on the carrier material contributes to an increase in the system's entropy. Based on entropy bond theory, compared to atomic aggregation, the uniformly dispersed state of atoms is more stable. Thus, metal atoms manifest as uniformly dispersed on the carrier, rather than forming nanoscale particles. Due to the presence of numerous structural defects in the N-doped carbon fibers (Figure S4, Supporting Information), metal atoms are effectively anchored onto these defects during diffusion, achieving an atomically dispersed state, thereby realizing the in-situ construction of SACs and DSACs.

Compared to the traditional "upstream-downstream" evaporation and capture strategies[36], the "integrated" in-situ construction approach we adopted demonstrates significant superiority, as depicted in the synthesis flowchart (Figure S5, Supporting Information). The approach demonstrates notable scalability, applicable to a wide range of metal precursors and support materials, thereby providing a robust synthetic route for diverse catalyst preparations. Of paramount significance is the precise control achieved over the metal loading ratio by finely tuning the metal salt proportions in the precursor solution. This meticulous control ensures the uniform distribution of metals in the catalyst and attains the desired metal content. The precision in metal loading serves as a pivotal tool for optimizing catalyst performance. Furthermore, the in-situ construction method streamlines the synthesis steps, enhancing overall synthetic efficiency while concurrently minimizing resource wastage and maximizing metal utilization. In our investigation, we employed a sandwich-like structure of electrospun membranes (multi-layer stacked membranes) to validate our intermediary strategy. Notably, electrospinning technology offers extensive potential applications, encompassing various structures beyond this design, including core-shell fiber membranes, mesh-type fiber membranes, composite fiber membranes, corrugated or folded fiber membranes, interwoven fiber membranes, and sponge-like fiber membranes, among others[37]. These diverse structural membrane materials, whether used individually or in combination, harbor significant innovative potential for supporting SACs and DSACs.

As shown in **Figure 1**b, we constructed a CuN$_4$-MN$_4$ (M=Cr, Mn, Fe, Co, Ni) DSACs model to elucidate the coordination environment and geometric configuration of the dual single atoms at the microscopic scale. This model is dedicated to offering detailed molecular insights into the electronic structure and localized characteristics of dual single atoms, emphasizing our dedication to the synthesis of DSACs.

Further, as illustrated in the Aberration-corrected HAADF-STEM image of CuNi DSACs (**Figure 1**c), we were able to partially validate the accuracy of the DSACs model. To further illustrate the versatility of our synthesis approach, we successfully fabricated five types of CuM (M=Cr, Mn, Fe, Co, Ni) DSACs. As depicted in **Figure 1**d, the low-magnification TEM images and corresponding EDS mapping images unequivocally demonstrate that these catalysts are atomically dispersed on the carbon fibers. The SEM images (Figure S6, Supporting Information) reveal no metal particles and successfully demonstrate the fabrication of carbon fibers with a diameter of approximately 100 nm. Furthermore, the XRD analyses (Figure S7, Supporting Information) provide additional confirmation, showcasing a lack of distinct crystalline peaks in the produced materials, underscoring the successful realization of DSACs.

For further comparison, we prepared nitrogen-doped carbon fibers (NC), Cu SACs, and Ni SACs. The XRD analyses (Figure S8,9, Supporting Information) showed no crystalline peaks, and the SEM images (Figure S4a,10, Supporting Information) indicated the successful fabrication of carbon fibers with a diameter of approximately 100nm. The low-magnification TEM images and corresponding EDS mapping images (Figure S11, Supporting Information) confirmed the atomic-level dispersion of metals on the carbon fibers, with Cu and Ni atoms uniformly distributed.

The combined metal content in CuNi DSACs amounts to 5.66 wt.%, comprising 3.00 wt.% Ni and 2.66 wt.% Cu, as determined by inductively coupled plasma optical emission spectroscopy (ICP-OES) (Table S1, Supporting Information). To further investigate the precision and feasibility of our adopted synthesis strategy in catalyst loading control, we intentionally adjusted the molar ratio of the metal precursors to Cu:Ni = 1:2. This step resulted in the highest final loadings of nickel and copper being 6.76 wt.% and 3.49 wt.%, respectively, totaling 10.25 wt.%. Furthermore, it's worth mentioning that the peak loading of Ni in the SACs achieved a notable 10.29 wt.%. (Table S2, Supporting Information)

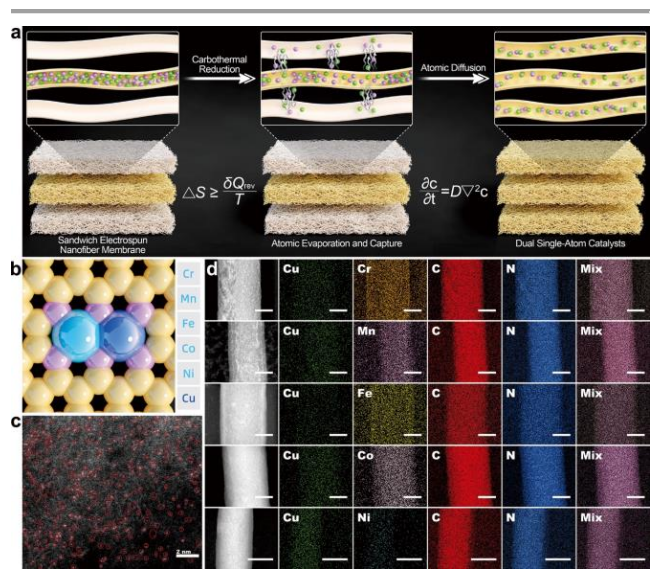

**Figure 1.** Scheme of EEMIS-DSAC Strategy. (a) Growth mechanism diagram for synthesizing DSACs. (b) Top-view structure of the CuM DSACs (M= Cr, Mn, Fe, Co and Ni). (c) Aberration-corrected HAADF-STEM image of CuNi DSACs. (d) low-magnification TEM images and the corresponding EDS mapping images of CuM DSACs (M= Cr, Mn, Fe, Co and Ni). Scale bar, 50 nm.

## 2 Characterization of DSACs

X-ray photoelectron spectroscopy (XPS) analyses (**Figure 2**a, S12, Supporting Information) elucidate the N 1s signal, revealing multiple nitrogen species. Specifically, the broad N 1s peak can be deconvoluted into pyridinic N at 398.5 eV, pyrrolic N at 400.9 eV, graphitic N at 402.7 eV, and Ni-N/ Cu-N at 399.7 eV[38, 39]. Single atoms also exhibit similar nitrogen species (Figure S13, Supporting Information). In the Cu 2p XPS spectra (Figure S14a, Supporting Information), the Cu 2p$_{3/2}$ signal displays two distinct peaks, attributed to the Cu$^+$ (932.81eV) and Cu$^{2+}$ (934.85eV), while the Cu 2p$_{1/2}$ signal displays two distinct peaks, attributed to the Cu$^+$ (952.50eV) and Cu$^{2+}$ (954.92eV)[40]. In the Ni 2p XPS spectra (Figure S14b, Supporting Information), the Ni 2p$_{3/2}$ peak resides between the values for Ni$^{2+}$ at 854.36 eV and 855.89 eV, while the Ni 2p$_{1/2}$ peak resides between the values for Ni$^{2+}$ at 871.67 eV and 873.41 eV[41, 42]. It's noteworthy that, in comparison to Cu SACs and Ni SACs, the binding energy of both the Cu 2p$_{3/2}$ peak and the Cu 2p$_{1/2}$ peak in CuNi DSACs exhibit a downward shift of 0.20 eV. Conversely, the Ni 2p$_{3/2}$ and Ni 2p$_{1/2}$ peaks in CuNi DSACs display an upward shift of 0.24 eV. The XPS peak shift of Ni/Cu dual atoms may imply electron transfer from Ni to Cu. The Raman spectrum (**Figure 2**b) reveals that the $I_D/I_G$ ratios for CuNi DSACs, Cu SACs, and Ni SACs are 1.23, 1.48, and 1.16, respectively. The findings suggest that CuNi DSACs exhibit a higher degree of structural defects compared to Cu SACs and a greater extent of graphitization relative to Ni SACs. Importantly, the incorporation of Ni significantly elevates the defect density in the carbon matrix, while the presence of Cu enhances the graphitization level. Notably, upon effective sequestration of copper ions, a majority of the carbon vacancies are occupied to form Ni-Cu pairs, resulting in a reduced $I_D/I_G$ ratio for CuNi DSACs[6].

Utilizing X-ray Absorption Fine Structure (XAFS) spectroscopy, we conducted an in-depth investigation into the local structure and coordination environment of Cu and Ni atomic sites in CuNi DSACs supported on Carbon Nanofibers (CNFs)[43]. **Figure 2**c presents the Cu K-edge X-ray absorption near-edge structure (XANES) spectra for CuNi DSACs, pure Cu foil, and Cu$_2$O. Intriguingly, the absorption edge of CuNi DSACs is situated between that of Cu$_2$O and CuO, revealing a mixed valence state of Cu atoms, encompassing both Cu$^{1+}$ and Cu$^{2+}$. Similarly, **Figure 2**d depicts the Ni K-edge XANES spectra for CuNi DSACs, indicating that the absorption edge lies between that of Ni foil and NiO. This suggests a mixed valence state for Ni atoms as well, specifically comprising Ni$^+$ and Ni$^{2+}$. Utilizing advanced Fourier Transform Extended X-ray Absorption Fine Structure (FT-EXAFS) analysis, we have elucidated the intricate coordination environments of Cu and Ni atomic sites in CuNi DSACs. In the Cu K-edge FT-EXAFS spectrum of CuNi DSACs (**Figure 2**e), a prominent peak at 1.50 Å is unequivocally attributed to Cu-N coordination. Significantly, the absence of the characteristic Cu-Cu bond at 2.24 Å corroborates the isolated nature of the Cu atoms. Similarly, in the Ni K-edge FT-EXAFS spectrum (**Figure 2**f), a distinct peak at 1.60 Å is ascribed to Ni-N coordination, and the lack of the typical Ni-Ni bond at 2.18 Å further substantiates the isolated state of Ni atoms. According to the quantitative FT-EXAFS results (Figure S15, S16 and Table S3, S4, Supporting Information), the average coordination numbers for Cu-N and Ni-N in CuNi DSACs are determined to be 3.9±0.1 and 4.2±0.6, respectively. This compelling evidence confirms the presence of Cu-N$_4$ and Ni-N$_4$ moieties in CuNi DSACs. Wavelet-transformed extended X-ray absorption fine structure (WT-EXAFS) analyses[44, 45], as depicted in **Figures 2**g-f, provide compelling evidence for the unique coordination environments of Cu and Ni atoms in CuNi DSACs. Intensity maxima observed at 6.2 Å$^{-1}$ and 4.0 Å$^{-1}$ is attributed to Cu-N and Ni-N, respectively, diverging significantly from the typical Cu-Cu and Ni-Ni contributions commonly found in pure Cu and Ni foils. Additionally, the absence of Cu-O and Ni-O, frequently encountered in Cu$_2$O and NiO (Figure S17, Supporting Information), further corroborates the distinctiveness of the CuNi DSACs. These findings not only confirm the presence of isolated Cu-N$_4$ and Ni-N$_4$ coordination sites in CuNi DSACs but also highlight their pronounced deviation from Cu SACs and Ni SACs. Particularly, energy shifts observed in XPS analyses suggest that the electronic structure alterations in CuNi DSACs are predominantly Engineered by bimetallic doping rather than nitrogen coordination. In summary, we conclusively identify isolated $Cu-N_4$ and $Ni-N_4$ coordination sites in CuNi DSACs, which engage in significant electronic interactions, thereby inducing substantial alterations in the electronic structure[39]. This conclusion is further substantiated by Aberration-corrected HAADF-STEM image of CuNi DSACs (**Figure 1**c) and is in excellent agreement with density functional theory (DFT) computational models.

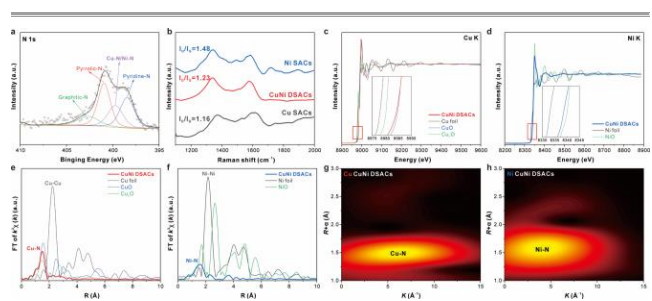

**Figure 2**. Atomic structural and chemical states analysis of CuNi DSACs. (a) High-resolution N 1s XPS spectrum of CuNi DSACs. (b) Raman spectra of Cu SACs, Ni SACs and CuNi DSACs. (c) Cu K-edge XANES spectra of CuNi DSACs, Cu foil, CuO, and $Cu_2O$. (d) Ni K-edge XANES spectra of CuNi DSACs, Ni foil and NiO. (e) $k^3$-weighted FT of $\chi(k)$-function from the Cu K-edge EXAFS of CuNi DSACs, Cu foil, CuO, and $Cu_2O$. (f) $k^3$-weighted FT of $\chi(k)$-function from the Ni K-edge EXAFS of CuNi DSACs, Ni foil and NiO. (g)WT images of the Cu K-edge from CuNi DSACs. (h) WT images of the Ni K-edge from CuNi DSACs.

## 3 Catalytic performance of DSACs for the NO3RR

As a proof-of-concept, to evaluate the potential of the as-obtained DSACs, CuNi DSACs was employed as an electrocatalyst for $NO_3RR$. Linear Sweep Voltammetry (LSV) measurements were conducted in a supporting electrolyte comprising 0.1 M $KNO_3$ and 1 M KOH. The electrocatalytic performance of CuNi DSACs was benchmarked against various control electrodes, including glassy carbon electrodes, NC, Cu SACs, and Ni SACs (**Figure 3**a). The glassy carbon electrode, which served as the catalyst support, manifested negligible current density variations, thereby confirming its electrochemical inertness. Similarly, NC, when devoid of metal loading, exhibited minimal changes in current density, underscoring the pivotal role of metal active sites in catalyzing $NO_3RR$. Upon increasing the applied potential, CuNi DSACs exhibited a marked enhancement in current density for nitrate reduction. Specifically, at −0.2 V vs. RHE, the current density reached 80 mA $cm^{-2}$, significantly outperforming Cu SACs (57 mA $cm^{-2}$) and Ni SACs (32 mA $cm^{-2}$). This compelling evidence strongly advocates for the superior efficiency of CuNi DSACs in facilitating $NO_3RR$.

LSV and *i-t* Curve tests in 1 M KOH, supplemented with varying nitrate ($NO_3^-$) concentrations (**Figure 3**b-c, S18-23, Supporting Information), revealed a pronounced sensitivity of CuNi DSACs towards $NO_3RR$. The reaction rate adhered to first-order kinetics and exhibited no significant change in the current density of the LSV curve when $NO_3^-$ concentrations exceeded 0.5 M, indicating that the active sites on the electrode surface are fully occupied by $NO_3^-$, reaching saturation. Therefore, further increasing the $NO_3^-$ concentration does not enhance the reaction rate. As the $NO_3^-$ concentration gradually increases, the Hydrogen Evolution Reaction (HER) reaction is inhibited to some extent, allowing for the maximum FE of $NH_3$ to be achieved at higher potentials, thereby resulting in a higher production rate of $NH_3$. However, as an inevitable consequence, the FE of $NO_2^-$, as a byproduct, also sees an increase. Under conditions with a $NO_3^-$ concentration of 2000 ppm(32.3mM), relevant to industrial wastewater[46, 47], we achieved a maximum FE for $NH_3$ of 96.97±0.11%, a mass productivity of 131.47±0.14 mg $h^{-1}$ $mg^{-1}$, and an area productivity of 10.06±0.01 mg $h^{-1}$ $cm^{-2}$(Catalyst unit area loading 0.0765 mg $cm^{-2}$). Under conditions with a $NO_3^-$ concentration of 0.1 M, we achieved a maximum FE for $NH_3$ of 94.09±0.52%, a mass productivity of 239.10±0.60 mg $h^{-1}$ $mg^{-1}$, and an area productivity of 18.30±0.05 mg $h^{-1}$ $cm^{-2}$. Under conditions with a $NO_3^-$ concentration of 0.25 M, we achieved a maximum FE for $NH_3$ of 94.56±0.46%, a mass productivity of 432.76±2.47 mg $h^{-1}$ $mg^{-1}$, and an area productivity of 33.12±0.19 mg $h^{-1}$ $cm^{-2}$. Under conditions with a $NO_3^-$ concentration of 0.5 M, we achieved a maximum FE for $NH_3$ of 90.61±0.16%, a mass productivity of 1024.50±0.99 mg $h^{-1}$ $mg^{-1}$, and an area productivity of 78.41±0.08 mg $h^{-1}$ $cm^{-2}$. Under conditions with a $NO_3^-$ concentration of 1M, we achieved a maximum FE for $NH_3$ of 91.71±0.66%, a mass productivity of 1023.88±7.35 mg $h^{-1}$ $mg^{-1}$, and an area productivity of 78.36±0.56 mg $h^{-1}$ $cm^{-2}$.

Electrocatalytic $NO_3RR$ is inherently complex, involving multi-electron processes and competing reactions. At lower potentials, $NO_2^-$ formation is favored, impeding the complete reduction to ammonia. Conversely, at elevated potentials, the competing HER becomes increasingly dominant, thereby undermining energy efficiency[48].

As delineated in **Figures 3**d-f, CuNi DSACs demonstrated superior FE and $NH_3$ production rate under the conditions of 1 M KOH + 0.1 M $KNO_3$, while maintaining a lower efficiency for $NO_2^-$ formation. This highlights the accentuated role of competitive HER at higher potentials. The unique architecture of CuNi DSACs, featuring dual single-atom sites, offers a rich landscape of active sites. Copper sites act as effective suppressors of HER and are adept at $NO_3^-$ adsorption and conversion. Conversely, nickel sites facilitate the subsequent reduction of $NO_2^-$ to $NH_3$. This atomic-level synergy not only optimizes the utilization of active sites but also significantly enhances reaction selectivity. The electrochemical active surface area (ECSA) of CuNi DSACs (**Figure 3**g) was rigorously evaluated using electrochemical double-layer capacitance measurements. Remarkably, the ECSA of CuNi DSACs substantially exceeded that of Cu SACs and Ni SACs, further corroborating its superior electrocatalytic performance. Subsequently, *i-t* Curve stability tests were conducted in a mixed electrolyte of 1 M KOH and 1 M $KNO_3$. As depicted in **Figures 3**h-i, the catalyst maintained a consistent current density of approximately 1 A $cm^{-2}$ across six 12-hour cycles, while sustaining FE for $NH_3$ close to 90%. Following 72 hours of $NO_3RR$, the structural integrity of the CuNi DSACs was substantiated through TEM and EDS mapping analyses, as evidenced in Figure S24, Supporting Information.

To unambiguously ascertain the origin of the synthesized $NH_3$ as deriving from $NO_3^-$, isotope labeling experiments were conducted employing both $^{14}NO_3^-$ and $^{15}NO_3^-$. Product identification and quantification were performed using $^1H$ NMR spectroscopy. Distinct splitting of the $^1H$ resonance into three symmetric signals was observed when $^{14}NO_3^-$ was used, confirming the formation of $^{14}NH_4^+$. Conversely, when $^{15}NO_3^-$ was utilized, the resonance split into two signals, corroborating the formation of $^{15}NH_4^+$ (Figure S25, Supporting Information). Additionally, to ensure the authenticity and reliability of the overall data, this experiment primarily employed the indophenol blue method (IBS) for $NH_3$ detection, supplemented by the Nessler's reagent method. Both methods demonstrated a high degree of consistency in the measured FE of $NH_3$ (Figure S26, Supporting Information).

Compared to previously reported catalysts (Table S5, Supporting Information), CuNi DSACs demonstrate superior yield and FE, highlighting the exceptional catalytic performance of the diatomic single-atom pairs. These findings underscore the industrial applicability and robustness of CuNi DSACs as a potent catalyst for $NO_3RR$.

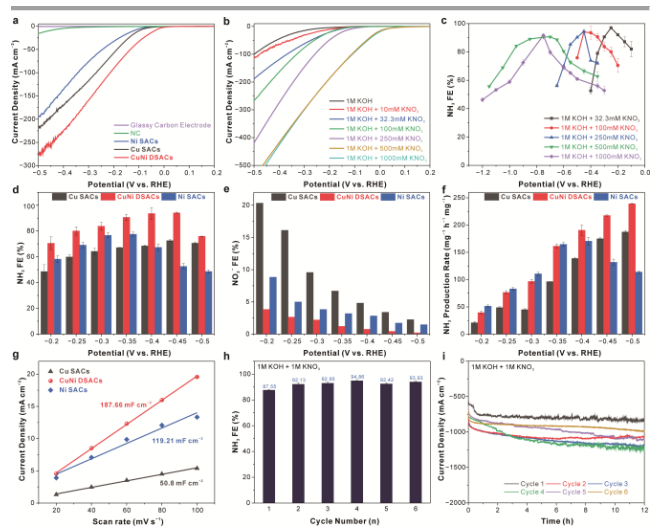

**Figure 3.** Electrocatalytic performance of CuNi DSACs for $NO_3RR$ (a) Linear sweep voltammograms of Glassy carbon electrode, NC, Cu SACs, Ni SACs and CuNi DSACs in a 1 M KOH with 0.1M $KNO_3$ electrolyte. (b) Linear sweep voltammograms of CuNi DSACs in 1M KOH containing varying concentrations of $KNO_3$. (c) FE of $NO_3RR$ to ammonia over CuNi DSACs in 1M KOH containing varying concentrations of $KNO_3$ at a series of potentials. (d) FE of $NO_3RR$ to $NH_3$ over Cu SACs, Ni SACs and CuNi DSACs in 1M KOH with 0.1M $KNO_3$ at a series of potentials. (e) FE of $NO_3RR$ to $NO_2^-$ over Cu SACs, Ni SACs and CuNi DSACs in 1M KOH with 0.1M $KNO_3$ at a series of potentials. (f) Production Rate of $NH_3$ over Cu SACs, Ni SACs and CuNi DSACs in 1M KOH with 0.1M $KNO_3$ at a series of potentials. (g) Double-layer capacitance per geometric area ($C_{dl}$) of Cu SACs, Ni SACs and CuNi DSACs. (h) FE of $NH_3$ in the cycling tests of $NO_3RR$ over CuNi DSACs at −0.75 V vs. RHE in 1M KOH with 1M $KNO_3$. (i) The i-t curves in the cycling tests of $NO_3RR$ over CuNi DSACs at −0.75 V vs. RHE in 1M KOH with 1M $KNO_3$.

## 4 Structural Analysis and Catalytic Activity of CuNi DSACs

To delve deeper into the structural and catalytic properties of CuNi DSACs, we designed a structure as depicted in **Figure 4**a. In this configuration, both Cu and Ni atoms are tetrahedrally coordinated with surrounding N atoms, bridged by the latter. Recognizing the significant electrostatic potential differences between these metal elements, we characterized this hypothesis by calculating the surface potential distribution of the material, as shown in **Figure 4**b. This reveals a marked potential difference between the Cu and Ni atoms. Such a difference induces an intrinsic electric field and generates surface-conjugated Lewis acid-base pairs, thereby augmenting the material's catalytic activity.

Additionally, this bonding configuration substantially boosts the spin charge density of the Ni atom, as demonstrated in **Figure 4**c. This enhancement is attributed to the dual single atom configuration, which alters the prevailing bond states in the system. This change results in a bonding structure that coexists in tetra-, penta-, and hexa-cyclic rings, effectively amplifying the coordination strength between the metal atoms and N (Figure S28, Supporting Information). Concurrently, the bridged N atoms, influenced by the charge from Cu/Ni elements, undergo polarization. This action deviates the metal coordination environment from the standard tetrahedral structure, which in turn modifies the arrangement of the metal element electrons in their respective orbitals. A comparison with the spin charge density of single atoms (Figure S29, Supporting Information) indicates a noticeable enhancement in the spin charge density of Ni in CuNi DSACs. The projected density of states (PDOS) in Figure 4d offers a clearer view of electron distribution across different energy levels. This shows a pronounced splitting near the Fermi level for Ni, and a similar phenomenon for Cu, suggesting both metals evolve as superior electron donors. Such properties enable a more effective bond with single-electron intermediates, consequently lowering the reaction energy barrier.

Building on these insights, we calculated the catalytic process of (2+6)-electron transfer from $NO_3$ to $NH_3$ (Figures 4e, S30). We observed that the rate-determining step for the single atom catalysts in this process is the conversion of *NO to *NOH (1.31 eV for Cu SACs and 1.40 eV for Ni SACs). In contrast, for CuNi SACs, the energy for this step is a mere 0.52 eV, with the entire reaction barrier being 0.79 eV. This can be attributed to the ability of the adsorbed structures during the reaction to rapidly toggle between the two metal sites, autonomously selecting the lowest energy configuration (Figure S31). Consequently, the catalytic activity of CuNi DSACs is notably superior to that of their single-atom counterparts, which aligns well with experimental outcomes.

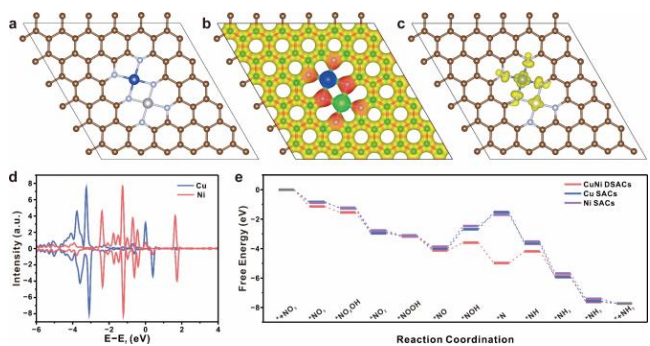

**Figure 4.** Structure and catalytic performance of CuNi DSACs. (a) Top view of the CuNi DSACs structure. (b) Surface potential density map. (c) Spin density distribution. (d) PDOS for distinct elements in the CuNi DSACs structure. (e) Reaction free energy profiles for various intermediates during the $NO_3$ reduction on CuNi DSACs at an applied potential of U = 0 V.

## CONCLUSIONS

In summary, compared to the traditional synthesis methods and strategies for SACs previously reported, the EEMIS-DSAC strategy we describe here offers a straightforward, precise, economical, and efficient avenue for large-scale production of DSACs. Integrating the conventional bottom-up and top-down approaches, we innovatively developed a synthesis method with broad adaptability, aimed at fabricating a variety of DSACs. Notably, the DSACs prepared through the EEMIS-DSAC strategy demonstrated outstanding industrial-level electrocatalytic performance in the NO$_3$RR, highlighting the immense potential of DSACs synthesized using the EEMIS-DSAC approach for future catalytic applications. Crucially, the EEMIS-DSAC strategy not only paves a promising path for the large-scale production of DSACs but also lays a solid foundation for utilizing DSACs to achieve sustainable production of high-value fuels and industrial raw materials.

## ASSOCIATED CONTENT

### Supporting Information

The Supporting Information is available free of charge on the ACS Publications website at http://pubs.acs.org.

> Experimental details; Morphological Characterization, X-ray Diffraction Analysis (XRD), Transmission Electron Microscopy (TEM), Scanning Transmission Electron Microscopy with Energy Dispersive X-ray Spectroscopy Mapping (STEM-EDS), High-Resolution X-ray Photoelectron Spectroscopy (XPS), Extended X-ray Absorption Fine Structure Analysis (EXAFS), Electrochemical Performance Testing, Isotopic Labeling for NH$_3$ Detection, Spin Density Mapping, and Schematic Representations of Catalytic Reaction.

## AUTHOR INFORMATION

### Corresponding Author

*du@jiangnan.edu.cn

### ORCID

Yao Hu: 0009-0005-2042-5895

Haihui Lan: 0000-0003-1356-4422

Mingliang Du: 0000-0003-2476-8594

### Author Contributions

§These authors contributed equally to this work

### Notes

The authors declare no competing financial interests.

## ACKNOWLEDGMENT

We acknowledge Zhang Qian from Shiyanjia Lab (www.shiyanjia.com) for the XPS analysis and the Jushang Scientific Research Service Platform (mall.jsceshi.cn) for providing XRD measurement.

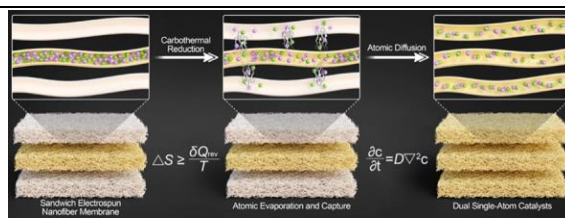

The study introduces the "Entropy-Engineered Middle-In Synthesis of Dual Single−Atom Compounds" (EEMIS-DSAC) strategy, a novel approach that seamlessly combines bottom-up and top-down synthesis paradigms. The method addresses challenges in DSAC synthesis, such as complexity and scalability. Notably, CuNi DSACs synthesized using EEMIS-DSAC displayed exceptional activity in nitrate reduction reactions, achieving impressive Faradaic efficiencies and productivities across varying nitrate concentrations. The innovative strategy paves the way for the fabrication of DSACs with broad adaptability and high performance.